\documentclass[twocolumn,amsmath,amssymb,pra]{revtex4-1}

\usepackage{amsmath} 

    \usepackage{graphicx}
    \usepackage{subfigure}
    \usepackage{color}
    \usepackage{dcolumn}
    \usepackage{bm}
    
    \usepackage{xspace}
    \usepackage{hyperref}

\newcommand{\SOM}[1]{\textcolor{red}{\cite{SOM}}}

    \newcommand{\ps}{phase space\xspace}

    \newcommand{\anon}[1]{#1}
   \newcommand{\figureLabel}[1]{(#1)}
\newcommand{\new}[1]{\textcolor{blue}{#1}}
\renewcommand{\new}[1]{#1}
\newcommand{\newnew}[1]{\textcolor{red}{#1}}
\renewcommand{\newnew}[1]{#1}

\begin{document}

\title{Anharmonic quantum mechanical systems do not feature \ps trajectories}

\author{Maxime Oliva, Dimitris Kakofengitis and Ole Steuernagel}

\address{School of Physics, Astronomy and Mathematics, University of Hertfordshire, Hatfield, AL10 9AB, UK 
}

\date{\today}

\begin{abstract}
  Phase space dynamics in classical mechanics is described by transport along
  trajectories. Anharmonic quantum mechanical systems do not allow for a trajectory-based
  description of their \ps dynamics. This invalidates some approaches to quantum \ps
  studies. We first demonstrate the absence of trajectories in general terms. We then give
  an explicit proof for all quantum \ps distributions with negative values: we show that
  the generation of coherences in anharmonic quantum mechanical systems is responsible for
  the occurrence of singularities in their \ps velocity fields, and vice versa. This
  explains numerical problems repeatedly reported in the literature, and provides deeper
  insight into the nature of quantum \ps dynamics.
\end{abstract}

\maketitle

\section{Introduction and Motivation}

The \ps dynamics of classical conservative mechanical systems is described by the
transport equations of Hamiltonian flow, along trajectories.  For quantum mechanical
systems this is only true for harmonic potentials, anharmonic quantum mechanical systems
do not transport quantum \ps distributions along trajectories.

This important fact is not appreciated by all: a number of incorrect schemes to model
quantum dynamics using \ps trajectories have been
devised~\cite{Lee_Scully_JCP82,Carruthers_RMP83,Lee_FoP83,Henriksen_CPL88,Lee_PLA90,Sala_JCP93,
  Muga_SSC95,Lee_PRep95,Razavy_PLA96,Dias_JMP02,Razavy_Book_03,Daligault_PRA03,Trahan_JCP03,
  Zhang_ChPL09,Brosens_SSC10,Sels_PASM12},
leading to
confusion~\cite{Wong_PRC82,Lee_JCP86,Skodje_PRA89,Wong_JOB03,Landauer_RMP94,Krivoruchenko_PRC07,Dittrich_Pachon_JCP10,Sels_PASM13}.
The schemes' failures have, in some quarters, given \ps representations of quantum
mechanics an undeservedly poor standing~\cite{PrivateCommunications_and_ReviewersReports}.

Here we revisit the basic features of quantum dynamics in
\ps~\cite{Schleich_01,Zachos_book_05,Hirshfeld_AJP02} in order to identify
concepts of classical dynamics that cannot be applied to quantum systems (recent reviews
on quantum-classical methods can be found
in~\cite{Miller_JCP12,Habershon_ARPC13,Koda_JCP15}). Our analysis deepens our
understanding of the behaviour of quantum dynamics in \ps. We show how the
generation of quantum coherences renders quantum dynamics in \ps very different
from classical dynamics.

In Section~\ref{sec:Classical_phase_space} we explain how \ps trajectories arise from
solutions of first order differential equations as integral curves describing the
transport of a density distribution.

In Section~\ref{sec:WhyWigner}, we emphasize that quantum \ps-based
studies~\citep[see][]{Schleich_01,Hirshfeld_AJP02,Zachos_book_05} of quantum dynamics are
no more involved than methods using von Neumann's equation to propagate the density
matrix.

Section~\ref{subsec:TrajConceptLure} emphasizes that a priori it is not clear whether
quantum dynamics can be described using trajectories.

In Section~\ref{sec_TimeEvol_W} we show that anharmonic systems are described by evolution
equations which are higher order differential equations, these generally do not permit a
trajectory description but create quantum coherences; we additionally emphasize that
harmonic systems cannot generate quantum coherences.

We explicitly prove in Section~\ref{sec_TimeEvol_Lagrangian} that quantum \ps
distributions with negative values (such as Wigner's distribution) cannot feature
trajectories because the quantum analog of Hamilton's \ps velocity field becomes
singular. We show why such singularities are \emph{needed} to create quantum coherences. 

The singularities affect numerical performance badly, see
reference~\cite{Moller_JPC94}, in Section~\ref{sec:Short_time_explosion}
we explain why, using a simple toy system.

Several misconceptions and incorrect conclusions drawn from ill-fated applications of the
trajectory concept are reported in the literature. Some are examined in
Section~\ref{App:Misconceptions} in order to explain how they fail and to further
illuminate differences between classical and quantum dynamics; before we conclude.

\section{Continuity equation, trajectories, classical \ps flow and Liouville's
  theorem \label{sec:Classical_phase_space}}
        
The transport of a density~$\rho(\bm{r},t)$, where the initial density~$\rho(\bm{r},0)$
and its current $\bm{j}$ encode the boundary conditions, is governed by a continuity
equation
\begin{eqnarray}\label{eq:Classical_Continuity}
\partial_t \rho(\bm{r},t) + \bm \nabla \cdot \bm{j}(\bm{r},t) = 0  \; .
\end{eqnarray}
Here we write~$\frac{\partial}{\partial t}=\partial_t$,
$\bm{r} = \genfrac{(}{)}{0pt}{}{\bm{x}}{\bm{p}}$
 parametrizes locations in phase space and
$\bm{\nabla} = \genfrac{(}{)}{0pt}{}{\partial_{\bm{x}}}{\partial_{\bm{p}}}$; we denote
vectors in bold face, and `$\cdot$' stands for scalar product.

A time-dependent solution for~$\rho$ of the Eulerian type (integrated over a
time-differential~$dt$ while keeping the position~$\bm{r}$ fixed), is of the form
 \begin{flalign}
\label{eq:euler_Solution}
\rho(\bm{r}, t+dt) = \; \rho(\bm{r},t) - dt \; \bm \nabla \cdot \bm{j} .
\end{flalign}
From a fluid dynamics perspective, the Eulerian approach tends to be in
\emph{conservation form} and its solutions therefore well behaved numerically.

\subsection{Trajectories through Lagrangian decomposition\label{subsec_LagrangeDecomp}}

If the current factorizes as
\begin{eqnarray}\label{eq:Classical_current}
    \bm{j}(\bm{r},t) = \rho(\bm{r},t) \bm{v}(\bm{r},t) ,
\end{eqnarray}
where~$\bm{v}$ is the velocity field, the continuity
equation~\eqref{eq:Classical_Continuity} can be rewritten in Lagrangian
decomposition~\cite{Donoso_PRL01,Trahan_JCP03,Daligault_PRA03}
\begin{flalign} 
\frac{d}{dt} \rho = \partial_t \rho + \bm{v} \cdot \bm \nabla \rho = - \rho \bm \nabla \cdot \bm{v} \; . 
\label{eq:Rho_LagrangianDecomposition}
\end{flalign}

If equation~(\ref{eq:Rho_LagrangianDecomposition}) is of first order in the derivatives,
linear in~$\rho$, and all quantities are mathematically well behaved, its solution allows
for a trajectory-based description, in Lagrangian (or co-moving) transport
form~\cite{Daligault_PRA03,Trahan_JCP03}
\begin{flalign} 
  \rho(\bm{r}_t, t) = e^{-\int_0^t d\tau \bm \nabla \cdot \bm{v}(\bm{r}_\tau,\tau) } \;
  \rho(\bm{r}_0, 0) \; ,
\label{eq:rho_Solution_Lagrangian}
\end{flalign}
where \emph{trajectories} are functions~$\bm{r}_t$, parameterized by time~$t$, arising
through integration of~$\bm{v}$ starting from~$\bm{r}_0$
\begin{flalign} 
  \bm{r}_t(\bm{r}_0) = \bm r_0 + \int_0^t d\tau \; \bm{v}(\bm{r}_\tau,\tau) .
\label{eq:Trajectory_Solution_Lagrangian}
\end{flalign}
Trajectories are integral curves describing the transport of a density distribution.

From a fluid dynamics perspective, the Lagrangian approach tends not to be in
conservation form and its solu\-tions therefore poorly behaved numerically.

From a mathematical perspective, solution~\eqref{eq:rho_Solution_Lagrangian} is found
using the method of characteristics (also known as Lagrange-Charpit method) which requires
the governing equation~\eqref{eq:Rho_LagrangianDecomposition} to be of first order in its
derivatives of~$\rho$. For example, diffusion equations are of second order and do not
admit trajectory-based solutions.

\subsection{Liouvillian flow in conservative classical mechanical
  systems\label{subsec_LiouvilleFlow}}

The natural setting for the dynamics of a mechanical particle is its phase
space~\cite{Nolte_PT10}. In this work we discuss a particle with mass~$M$ moving in one
dimension~$x$ only. The associated two-dimensional \ps is parameter\-ised by
vectors $\bm{r} = \genfrac{(}{)}{0pt}{}{x}{p}$, subject to a conservative hamiltonian
$H=p^2/(2M)+V(x)$. In this case, the particle's \ps velocity
\begin{eqnarray}\label{eq:Classical_flow}
\bm
v(\bm{r}) = \frac{d}{dt}\bm{r}_t = \genfrac{(}{)}{0pt}{}{{p}/{M}}{-\partial_x V(x)} \; ,
\end{eqnarray}
encapsulates Newton's laws, and features volume preserving or `Liouvil\-lian'
dynamics:~$\bm \nabla \cdot {\bm{v}}=0$.  As a function of~$\bm{r}$ only, $\bm{v}$~is
independent of time~$t$ and state~$\rho(\bm{r},t)$.

The Liouvillian nature of classical Hamiltonian current implies with
$\bm \nabla \cdot \bm{j} = \bm{v} \cdot \bm \nabla \rho $, that the total
derivative~\eqref{eq:Rho_LagrangianDecomposition} is zero: the value of $\rho$, while the
dynamics sweeps it along its trajectories, stays
constant~$\rho(\bm{r}_t(\bm{r}_0),t)=\rho(\bm r_0,0).$ In this case,
solution~\eqref{eq:rho_Solution_Lagrangian}, viewed as the function~$\rho(\bm{r},t)$,
through relabelling $\bm{r} = \bm{r}_t $, simplifies to the pull-back form
\begin{eqnarray}\label{eq:rho_TimeEvolution}
  \rho(\bm{r},t) = \rho\left(\bm{r} - \int_0^{t} d\tau \; \bm{v}(\bm{r}_\tau),0\right) \; .
\end{eqnarray}

\subsection{A simple system with non-Liouvillian flow that features
  trajectories\label{sec_FrictionViolatesLiouville}}

A free particle slowed down by friction $\dot p = -\gamma p$ is a classical system
violating Liouville's theorem. With $p_t=p_0 \; e^{-\gamma t}$ and
$x_t=x_0 + (1-e^{-\gamma t}) p_0/(m\gamma)$, Kramer's evolution equation
$\partial_t \rho = [-p/m \partial_x + \gamma p \partial_p + \gamma ] \rho$, where the
diffusive Brownian motion term has been neglected, yields the trajectories-based solution
of transport form~\eqref{eq:rho_Solution_Lagrangian}
\begin{flalign}
\label{eq:rho_Solution_ClassicalFriction}
\rho(x,p,t) = \exp[\gamma t] \; \rho_0(x-\frac{p}{\gamma m}(e^{\gamma t} -1), p \; e^{\gamma t}) \; .
\end{flalign}
The coefficient function $\exp[\gamma t] $ keeps this distribution normalized while the
dynamics shrinks volumes uniformly across \ps: $ \bm{\nabla} \cdot \bm v =
\bm{\nabla} \cdot (p/m , -\gamma p) = -\gamma$.

\section{Wigner's quantum \ps distribution \label{sec:WhyWigner}}

Attempts to understand and numerically approximate quantum dynamics of anharmonic systems
has frequently relied on the concept of \ps trajectories in a way unsuitable for
this
task~\cite{Lee_Scully_JCP82,Carruthers_RMP83,Lee_FoP83,Henriksen_CPL88,Lee_PLA90,Sala_JCP93,
  Muga_SSC95,Lee_PRep95,Razavy_PLA96,Dias_JMP02,Razavy_Book_03,Daligault_PRA03,Trahan_JCP03,
  Zhang_ChPL09,Brosens_SSC10,Sels_PASM12}.
This seems to be the reason for the fatigue expressed by fellow researchers who perceive
the `Wigner method' (and other \ps methods) as unsuitable for finding ways of
reducing computational complexities~\cite{PrivateCommunications_and_ReviewersReports}.
  
  We have not found a rigorous explanation for the supposed unsuitability of Wigner's
  representation of quantum mechanics. In the general case it is not justifiable since
  \emph{``All calculation methods scale in proportion to the volume of phase space that
    the molecular encounter occupies. Therefore, phase space is a common denominator by
    which different methods of calculation can be compared and the feasibility of the
    calculation estimated.''}~\cite{Kosloff_JPC88}

Recent work shows that the propagation of the Wigner distribution is suitable for the
study of quantum dynamics of anharmonic systems~\cite{Cabrera_PRA15} and that its study
provides new valuable insight~\cite{Koda_JCP15}.

A quantum state's density matrix $ \varrho(x,x',t) = \langle x|\hat \varrho(t) |x'\rangle$,
can equivalently be described by Wigner's \ps-based quantum
distribution~$W(x,p,t)$~\cite{Wigner_PR32,Hillery_PR84,Zachos_book_05,Case_AJP08}, both
are based in spaces of equal dimension:
\begin{equation}\label{eq:W}
  W_\varrho(x,p,t) \equiv \frac{1}{\pi \hbar} \int_{-\infty}^{\infty}
 dy \, e^{-\frac{2i}{\hbar} p y} 
 \langle x+y| \hat \varrho(t) | x-y \rangle  \; .
\end{equation}

$W$ can numerically be generated quickly through fast Fourier transforms of~$\varrho$.

$W$ is real-valued (unlike~$\varrho$), non-local (through~$y$), and normalized 

$ \int_{-\infty}^{\infty} \int_{-\infty}^{\infty} dx \; dp \; W(x,p,t) = 1 $.

Generally, the Wigner distribution has negative patches~\cite{Hudson_RMP74}, like many
other quantum \ps distributions~\cite{Hillery_PR84}, this will be important for
part of our discussion, in Section~\ref{sec_TimeEvol_Lagrangian}.

Specifically, Wigner's distribution is set apart from other quantum \ps
distributions~\cite{Hillery_PR84} by the fact that only Wigner's simultaneously yields the
correct projections in Schr\"odinger's position
$\varrho(x,x,t)= \int_{-\infty}^{\infty} dp \; W(x,p,t)$ and momentum representation
$\tilde \varrho(p,p,t)=\int_{-\infty}^{\infty} dx \; W(x,p,t)$, while maintaining its
form~(\ref{eq:W}) when evolved in time and giving the overlap between states in the simple
form~$|\langle \psi_a | \psi_b \rangle |^2 = 2 \pi \hbar \int_{-\infty}^{\infty} dx
\int_{-\infty}^{\infty} dp \; W_a \; W_b$.  Finally, the Wigner distribution's averages
and uncertainties evolve momentarily
classically~\cite{Royer_FOP92,Ballentine_PRA94}. $W$~is considered the \emph{``closest
  quantum analogue of the classical phase-space distribution''}~\cite{Zurek_NAT01}.

For specificity we choose Wigner's distribution for our discussions of quantum \ps
behaviour. Most of our results apply to other quantum \ps distributions as well; the proof
in section~\ref{sec_TimeEvol_Lagrangian} explicitly applies only to
those~\cite{Hillery_PR84} that have negative patches in \ps.

\section{Trajectories in quantum systems\label{subsec:TrajConceptLure}}

Note that by trajectories we mean integral curves that obey the equations of motion, we
neither discuss paths (which do not have to follow equations of motion) nor center-of-mass
trajectories as discussed by Heisenberg~\cite{Heisenberg_ZfP27}.

A priori it is not clear that one must not use trajectories for quantum \ps
descriptions of anharmonic systems.

Heisenberg's uncertainty principle is at times interpreted to mean that quantum mechanics
does not allow for a trajectory-based description. This interpretation is incorrect:

Phase space trajectories are a fruitful mathematical device for the description of quantum
dynamics of a system if the potential is of the quadratic
form~(\ref{eq:H_quadratic}). This statement applies to non-dissipative systems, even
driven ones. For such systems the trajectory-description~\eqref{eq:rho_TimeEvolution}
(for~$W$ rather than $\rho$) applies. Using trajectories, which in this case follow the
classical law~\eqref{eq:Classical_flow}, is in fact simpler and, in this sense, even
superior to the use of standard Schr\"odinger wave function propagators, see Takabayasi in
Ref.~\cite{Takabayasi_PTP54}~p.~352.

Bohm's representation of quantum theory uses configu\-ration space
trajectories~\cite{Lopreore_PRL99,Hiley_Proc04} and these have experimental
relevance~\cite{Kocsis_SCI11}.

The concept of paths has been fruitful in path-integral formalisms applied to
configuration or phase space.

Semiclassical methods employ classical trajectories along which quantum objects are
carried~\cite{Heller_JCP76,Tomsovic_Heller_PRL91,Berry_JPA79}.

When trajectory techniques can be implemented for quantum dynamical \ps studies
they permit us to launch large numbers of trajectories while allowing us to efficiently
parallelize computer code~\cite{Moller_JPC94,Trahan_JCP03}.

In what follows, we will, however, see that in anharmonic quantum systems the divergence
of the velocity field in \ps is non-zero. One might still hope to describe the
propagation of~$W$ in \ps by Eq.~\eqref{eq:rho_Solution_Lagrangian} [or
Eq.~\eqref{eq:W_Solution_Integrated}]. But it turns out that the divergence of the quantum
mechanical velocity field in \ps is singular, see
Section~\ref{sec_TimeEvol_Lagrangian} and Fig.~\ref{fig:divergence}~(c). This cannot be
avoided~\cite{Kakofengitis_PRA17}, and therefore we can explicitly prove, by
contradiction, that trajectories do not exist globally for systems whose \ps
distributions can develop areas with negative values, see
Section~\ref{sec_TimeEvol_Lagrangian}.

\section{Time evolution of the Wigner distribution\label{sec_TimeEvol_W}}

The time evolution of~ $W(x,p,t)= W(\bm{r},t)$ is given by the Eulerian continuity
equation~\cite{Wigner_PR32}
\begin{flalign} 
  \partial_t W(\bm{r},t) = - \bm \nabla \cdot {\bm{J}(\bm{r},t)} \; .
\label{eq:W_Continuity}
\end{flalign}
Generally, the Wigner current~$\bm{J}$ has an integral
representation~\cite{Wigner_PR32,Hillery_PR84,Baker_PR58,Kakofengitis_PRA17}, but for
potentials~$V(x)$ that can be Taylor-expanded, giving rise to finite forces only, $\bm{J}$
is of the Moyal-form~\cite{Wigner_PR32,Moyal_MPCPS49}
\begin{eqnarray} {\bm{J}} = \binom{ J_x }{ J_p } = \bm{j} +
 \begin{pmatrix} 0
    \\-\sum\limits_{l=1}^{\infty}{\frac{(i\hbar/2)^{2l}}{(2l+1)!}
      \partial_p^{2l}W \partial_x^{2l+1}V }
 \end{pmatrix}\!.
\label{eq:FlowComponents}
\end{eqnarray}
Here, with $\bm{j} = W \bm{v}$, $\bm{J} - \bm{j}$ are the `quantum-correction' terms.

Fieldlines of Wigner current are well defined and their depiction has helped to reveal the
topological charge conservation of~$\bm{J}$'s stagnation
points~\cite{Ole_PRL13,Kakofengitis_EPJP17}.

In analogy to the classical Euler solution~\eqref{eq:euler_Solution}, the integration of
the continuity equation~(\ref{eq:W_Continuity}) yields
\begin{flalign}
\label{eq:W_better_dt}
W(\bm{r}, t+dt) = \; W(\bm{r},t) - dt \; \bm \nabla \cdot \bm{J},
\end{flalign}
which is in conservation form.

\subsection{Formation of coherences and negativities of the Wigner
  distribution\label{subsec_CoherencesNegativities}}

The primary difference between classical and quantum states is the ability of a quantum
particle to form non-local coherences (to be present in both holes of a double
slit~\cite{Feynman_NegEssay87,Heller_JCP76}). Precisely these nonlocal coherences in
configuration space are revealed by the Wigner distribution's negative
patches~\cite{Feynman_NegEssay87} in \ps, and vice
versa~\cite{Leibfried_PT98,Zurek_NAT01,Grangier_SCI11}. Coherences or negative patches can
only be generated in anharmonic systems. Harmonic systems (and their isomorphic
partners~\cite{Ole_FreeHOSC_14}) are of the quadratic form
\begin{flalign}
\label{eq:H_quadratic}
\hat H_{quadratic} (\hat x, \hat p) = \frac{\hat p^2}{2M} + \frac{K}{2} \hat x^2 + a \hat x + b
\end{flalign}
(here $K$, $a$ and $b$ are any real constants). They can feature
negative patches of the Wigner distribution only if these are inserted into the initial
condition,~$W_0(\bm{r})$, but they cannot \emph{generate} them, see
Eq.~\eqref{eq:W_Solution_Integrated} below.

\section{Singularities in the velocity field are needed: trajectories are
  ill-defined\label{sec_TimeEvol_Lagrangian}}

We now prove that the Lagrangian transport form is ill-defined in the quantum case.
Following references~\cite{Donoso_PRL01,Trahan_JCP03,Daligault_PRA03} we rewrite
continuity equation~(\ref{eq:W_Continuity}) for~$W$ in Lagrangian
decomposition~\eqref{eq:Rho_LagrangianDecomposition}
\begin{equation} 
  \frac{d W}{d t} = \partial_t W + \bm{w} \cdot \bm \nabla W =- W \bm \nabla \cdot \bm{w} \; . 
\label{eq:W_TotalDeriv}
\end{equation}
Here, the quantum \ps velocity field~$\bm
w$~\cite{Donoso_PRL01,Trahan_JCP03,Daligault_PRA03}, corresponding to the hamiltonian
velocity field~$\bm{v}$, is
\begin{equation}
\bm{w} = \frac{\bm{J}}{W} =\bm{v} + \frac{1}{W}
 \begin{pmatrix} 0
    \\-\sum\limits_{l=1}^{\infty}{\frac{(i\hbar/2)^{2l}}{(2l+1)!}
      \partial_p^{2l}W \partial_x^{2l+1}V } 
 \end{pmatrix} \; .
 \label{eq:Wigner_w_velocity}
\end{equation}

$\bm{w}$ is singular at zeros of~$W$ since, generally, zeros of~$W$ do not coincide with
zeros of its derivatives~\cite{Kakofengitis_EPJP17}. 

For time-differentials~$dt$, the formal solution of Eq.~(\ref{eq:W_TotalDeriv}), written
in pull-back form, like Eq.~(\ref{eq:rho_TimeEvolution}), has the transport
form~\cite{Daligault_PRA03,Trahan_JCP03}
\begin{flalign}
\label{eq:W_Solution_dt}
  W(\bm{r}, t+dt) =   e^{-dt \bm \nabla \cdot \bm{w}} \;  W(\bm{r} - dt\; \bm{w} ,t) 
\; ,
\end{flalign}
where the transport shift can be expressed via a translation using the convective
operator~$\bm{w} \cdot \bm \nabla $
\begin{flalign} 
 W(\bm{r} - dt\; \bm{w} ,t) =  e^{-dt \; \bm{w} \cdot \bm \nabla } \left[  W(\bm{r},t) \right]  \; . 
 \label{eq:dWR}
\end{flalign}

We emphasize that the Lagrangian decomposition, although technically correct, splits up
the well behaved expressions in the continuity equation~(\ref{eq:W_Continuity}) and in
this way creates singularities in the evolution equation~(\ref{eq:W_TotalDeriv}) and the
exponents of its solution~(\ref{eq:W_Solution_dt}) and~(\ref{eq:dWR}).

Following references~\cite{Daligault_PRA03,Trahan_JCP03} we formally extend the
integration in time for the transport form~(\ref{eq:W_Solution_dt}). To this end we
temporarily assume that globally~$W>0$ in order to remove singularities in~$\bm w$ in
Eq.~(\ref{eq:Wigner_w_velocity}) and that, additionally, $W$ and $V$ are of such a form
that~$|\bm \nabla \cdot \bm{w}|<\infty$. We formally arrive at the integrated transport
form~\cite{Daligault_PRA03,Trahan_JCP03}
\begin{flalign} 
  W(\bm{r}_t, t) = e^{-\int_0^t d\tau \bm \nabla \cdot \bm{w}(\bm{r}_\tau,\tau) } \;
  W(\bm{r}_0, 0) \; .
\label{eq:W_Solution_Integrated}
\end{flalign}
Trahan and Wyatt deduced~\cite{Trahan_JCP03} ``\emph{two important non-crossing rules that
  follow directly from Eq.~[(\ref{eq:W_Solution_Integrated})]: (i) a trajectory cannot
  cross a surface on which the density is zero; (ii) the sign of the density riding along
  the trajectory cannot change.}''~\cite{Trahan_Wyatt_(i)_comment}

Hudson's theorem~\cite{Hudson_RMP74}, however, shows that for anharmonic systems the
Wigner distribution can at best be positive everywhere in \ps for one point in
time only. Time evolution in an anharmonic potential immediately introduces zero-lines
of~$W$ somewhere in \ps. At~$W$'s zeros the Picard-Lindel\"of theorem is
violat\-ed, integrals~\eqref{eq:Trajectory_Solution_Lagrangian},~(\ref{eq:W_Solution_dt}),
(\ref{eq:dWR}) and~(\ref{eq:W_Solution_Integrated}) do not exist globally. Therefore, the
Lagrangian transport solution and trajectories~$\bm{r}_t$ do not exist across \ps.

In other words, Eq.~(\ref{eq:W_Solution_Integrated}) proves that bounded magnitude of
$\bm \nabla \cdot \bm{w}$ precludes sign changes in~$W$ along streamlines
of~$\bm{w}$.

This leads to one of our central results: \emph{the singularities of~$\bm{w}$ 
are {needed to create the negativities of~$W$}, i.e., they are needed to create quantum
coherences} (Section~\ref{subsec_CoherencesNegativities}).

\subsection{The phase space velocity~\texorpdfstring{$\bm{w}$}{} is non-linear in $W$ \label{subsec_nonLinear_w}}

In general, an evolution equation with higher order derivatives does not allow for
trajectory-based solutions since it is neither of first order in derivatives nor linear
in~$W$, see Section~\ref{subsec_LagrangeDecomp}. Forcing such an equation into Lagrangian
decomposition~\eqref{eq:W_TotalDeriv} leads to equations~\eqref{eq:Wigner_w_velocity}
which burden us with spurious non-linearities in~$W$: while
$\bm{J}(W_a) + \bm{J}(W_b) = \bm{J}(W_a + W_b)$, in general
$\bm{w}(W_a) + \bm{w}(W_b) \neq \bm{w}(W_a + W_b)$ and left and right hand side of
Eq.~\eqref{eq:W_TotalDeriv} are also non-linear in~$W$. This argument carries over to
evolution equations~\cite{Takahashi_JPSJ86,Veronez_JPA13} of other quantum \ps
distributions.

\section{Short time integration of Eulerian and Lagrangian evolution equations --an
  analytically solvable case-- \label{sec:Short_time_explosion}}

We now demonstrate that application of the Lagrangian
decomposition~\eqref{eq:W_TotalDeriv} creates misleading analytical and numerical results,
whereas the Eulerian approach gives correct results.

\begin{figure}[t]
   \includegraphics[width=0.49\columnwidth]{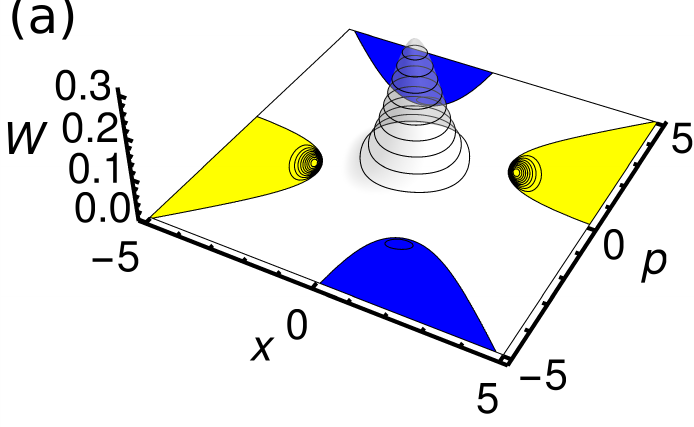}
  \includegraphics[width=0.49\columnwidth]{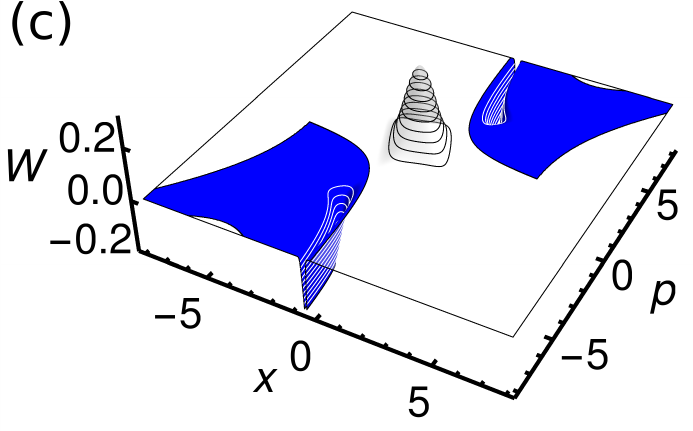}
\\
  \includegraphics[width=0.49\columnwidth]{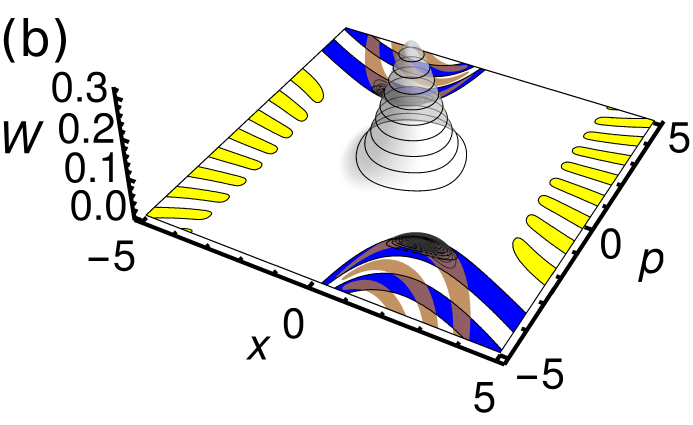}
  \includegraphics[width=0.49\columnwidth]{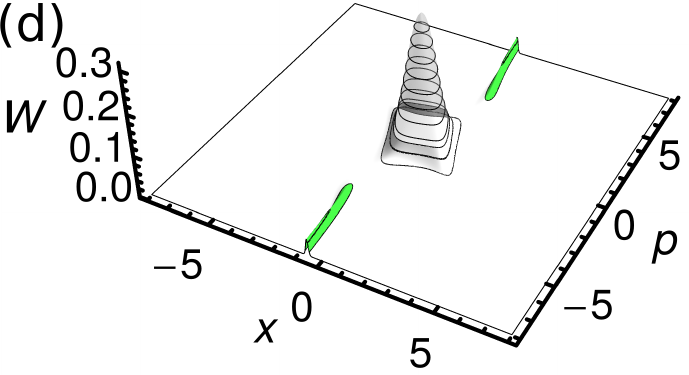}
  \caption{{Initially positive Wigner distribution~$W$ evolved in Eulerian
      form~\figureLabel{a} and~\figureLabel{b} or Lagrangian form~\figureLabel{c}
      and~\figureLabel{d}}. \figureLabel{a}, the single step propagation of $W_0$
    with~$\hbar=1, M=1, V= x^4,$ and $\Delta t = 10^{-2}$, using the Eulerian
    solution~(\ref{eq:W_Evolved_PathForm_1_Step}), shows correct formation of negativities
    as blue patches in \ps. However, due to the crude nature of the first order difference
    approximation~(\ref{eq:W_Evolved_PathForm_1_Step_A}) of the $\Delta t$-step employed
    in this illustration, spurious negative classical transport patches form, shown in
    yellow.  \figureLabel{c}, essentially the same scenario as~\figureLabel{a} (for
    explicitness we chose a longer time of~$\Delta t = 5 \times 10^{-2}$) is displayed
    using the Lagrangian transport form~(\ref{eq:W_x4_Transport_Form}): deep, unphysical
    gashes form due to the singularities in~$\bm{w}$
    in~(\ref{eq:w_x4_Transport_Form}). \figureLabel{b}, using a twelve step iteration of
    the Eulerian solution~(\ref{eq:W_Evolved_PathForm_1_Step}), while reducing the time
    step per iteration to $10^{-2}/12$, we end up with a better approximation than in~\figureLabel{a}: $W$'s
    negativities (blue patches) persist and develop fringes, whereas the unphysical
    (yellow) classical patches recede. This is confirmed by an exact numerical integration
    (brown overlay). \figureLabel{d}, the transport shift form~(\ref{eq:dWR}), for the
    same scenario as~\figureLabel{c}, displays unphysical formation of humps highlighted
    in green, their positions confirm that the singularities of~$\bm{w}$ create the deep
    gashes in~\figureLabel{c}.
    \label{fig:Negativty_from_ContinuityEq}}
\end{figure}

\new{ We use the harmonic oscillator
  groundstate~$W_0(\bm{r})=(\hbar \pi)^{-1}\exp[-(x^2+p^2/\hbar^2)]$ as a globally
  positive initial state for a quartic oscillator~$V(x) = K x^4$, $K>0$. Since it is not
  an energy eigenstate and all states have at least one zero at infinity, it immediately
  develops negativities~\cite{Hudson_RMP74}.
  \\
  \indent We find that even in this case, where initially singularities of $ \bm{w}$ and
  $\bm \nabla \cdot \bm{w}$ are absent, the Lagrangian approach fails, see
  Fig.~\ref{fig:Negativty_from_ContinuityEq}.} 

A first order difference approximation of continuity equation~(\ref{eq:W_Continuity}),
using a finite value for~$\Delta t$, applied to~$W_0(\bm{r})$, according to the Eulerian
equation~(\ref{eq:W_better_dt}) gives the single-step propagation approximation
\begin{flalign} \label{eq:W_Evolved_PathForm_1_Step_A}
  W&(\bm{r},\Delta t) \approx W_0(\bm{r})\; -\Delta t \; \; \bm \nabla \cdot \bm{J} \\
   & = \left[ 1 + \Delta t \left(-\hbar^2 K x \partial_p^3 + \{ 4 K x^3 \partial_p -
      \frac{p}{M} \partial_x \} \right) \right]\; W_0(\bm{r}) \; .
\label{eq:W_Evolved_PathForm_1_Step}
\end{flalign}

Eq.~(\ref{eq:W_Evolved_PathForm_1_Step}) is illustrated in
Fig.~\ref{fig:Negativty_from_ContinuityEq} \figureLabel{a}. It confirms the immediate
formation of negativities and that even the single-step approximation of the Eulerian
equation~(\ref{eq:W_better_dt}) gives tolerable results.  Repeated iteration of
Eq.~(\ref{eq:W_Evolved_PathForm_1_Step}) yields successively better approximations of the
true dynamics, see Fig.~\ref{fig:Negativty_from_ContinuityEq} \figureLabel{b}.

We now show that the growth of the magnitude~$|\bm \nabla \cdot \bm{w}|$, at small values
of~$W$ (even if~$W_0>0$ everywhere), is so explosive that it renders the Lagrangian
approach misleading, for any non-zero timestep~$\Delta t$.  We repeat
calculation~(\ref{eq:W_Evolved_PathForm_1_Step}) using the transport
form~(\ref{eq:W_Solution_dt}). In this case, for the same initial state~$W_0$, we get
the Lagrangian single-step propagation approximation
\begin{flalign} \label{eq:W_x4_Transport_Form} W(\bm{r}, \Delta t) \approx W_0(\bm{r}
  -\Delta t \; \bm{w}) \; [1 - \Delta t \; 8 \hbar^{-2} K\; x \; p] \; ,
  \\
\label{eq:w_x4_Transport_Form}
\text{where   } \bm{w} = 
\begin{pmatrix} {p}/{M} \\- 4 K x^3 + {\hbar^2} K x \; {W^{-1}} \;
      \partial_p^{2}W 
 \end{pmatrix}
\; .
\end{flalign}

Equation~(\ref{eq:W_x4_Transport_Form}) is incorrect, see
Fig.~\ref{fig:Negativty_from_ContinuityEq}~\figureLabel{c}, it puts the Wigner
distribution's negative patches into the wrong sectors in \ps and creates deep gashes in
them, see Fig.~\ref{fig:Negativty_from_ContinuityEq}~\figureLabel{d}; these violate
probability conservation.

We cross-checked these results, see
Fig.~\ref{fig:Negativty_from_ContinuityEq}~\figureLabel{b}, using standard numerical
Schr\"odinger function solvers, the `QuTiP' programming suite~\cite{QTip_CPC13} and a
split-operator technique~\cite{Cabrera_PRA15}, confirming that only the Eulerian
equation~(\ref{eq:W_Evolved_PathForm_1_Step}), see
Fig.~\ref{fig:Negativty_from_ContinuityEq}~\figureLabel{a}, and iterations thereof, see
Fig.~\ref{fig:Negativty_from_ContinuityEq}~\figureLabel{b}, give acceptable results.

Fig.~\ref{fig:Negativty_from_ContinuityEq}~\figureLabel{d} shows that even for our
initially positive state~$W_0$, the behaviour of~$\bm{w}$ is responsible for the stark
deviation of the ill-defined Lagrangian transport form~(\ref{eq:W_Solution_dt}) from the
correct Eulerian continuity equation's solution.

\section{Misconceptions associated with phase space
  trajectories\label{App:Misconceptions}}
\subsection{There are no Wigner trajectories\label{App:No_Wigner_Trajectories}}

\begin{figure*}[t]
\centering
   \includegraphics[width=1.465\columnwidth]{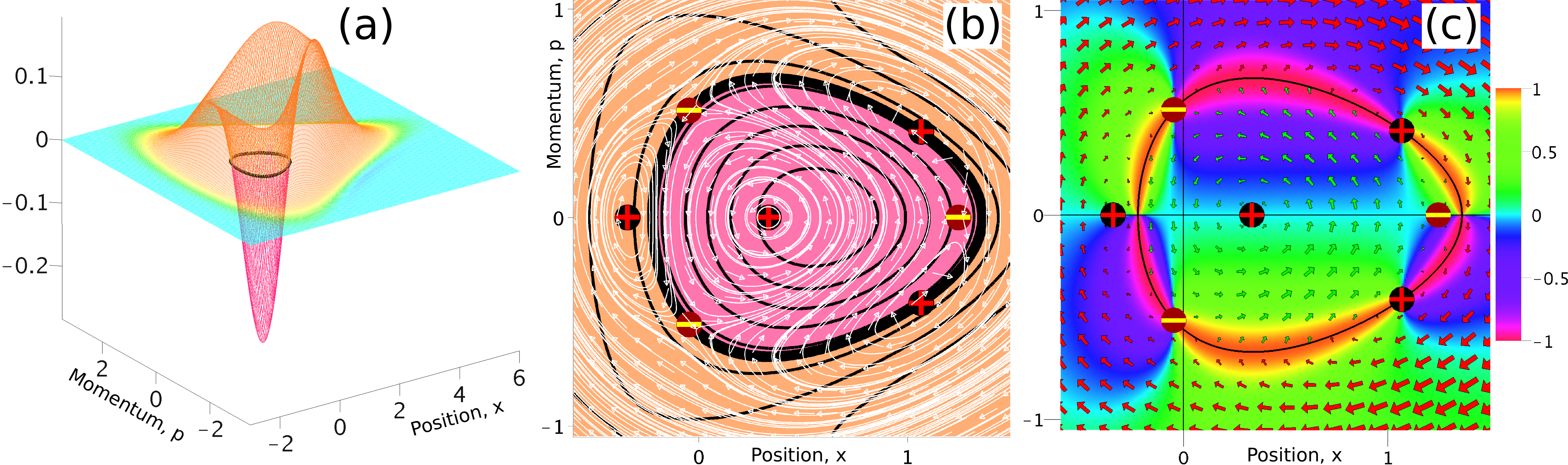}
   \caption{\figureLabel{a} The Wigner distribution for the first excited state of
     an anharmonic Morse oscillator~\cite{Dahl_JCP88} with
     potential~$U(x)=3(1-\exp({{-x}/{\sqrt{6}}}))^2$ (parameters:~$\hbar=1$ and $M=1$) is
     depicted by its black contour lines in~\figureLabel{b}. $W$'s thick zero
     contour (black line in~\figureLabel{a}) separates the negative central patch
     from the surrounding positive area in~\figureLabel{b}: Red crosses and yellow
     bars mark the locations of $\bm{J}$'s stagnation points, with Poincar\'e-Hopf
     indices~\cite{Ole_PRL13}~$\omega=+1$ and~$-1$, respectively. Integrated fieldlines
     of~$\bm{J}$ are depicted as thin white lines, displayed together with normalized
     current ${\bm{J}}/J$ (white arrows). $\bm{J}$-fieldlines, cut across the Wigner
     distribution's contours and enter and leave the negative area. \figureLabel{c}
     shows $\frac{2}{\pi} \arctan |\bm \nabla \cdot \bm{w}|$ and illustrates that $\bm{w}$
     is Liouvillian only on lines in \ps (cyan coloring) while featuring singular
     behaviour where $W=0$ (thin black line). Red arrows depict regular and green arrows
     inverted~\cite{Ole_PRL13} current~$\bm{J}$.
    \label{fig:divergence}}
\end{figure*}

In reference~\cite{Lee_Scully_JCP82} Lee and Scully consider energy eigenstates (of the Morse
potential) and argue that ``\emph{In terms of the Wigner distribution, it means that each
  \ps point should move in such a way that the Wigner distribution does
  not change in time. This consideration leads to the concept of `Wigner trajectories',
  trajectories along which \ps points of the Wigner distribution
  move. For the case under consideration, Wigner trajectories must be trajectories along
  the surfaces on which the Wigner distribution takes on the same value, i.~e.,
  trajectories along the equi-Wigner surfaces. These Wigner trajectories are
  `quantum-mechanical' trajectories in the sense that they represent paths of \ps
  points that move according to the quantum-mechanical equation of motion. They describe
  the exact quantum-mechanical dynamics in a \ps, whereas classical trajectories
  obviously yield only an approximate description of quantum
  dynamics}''\cite{Lee_Scully_JCP82}.

It has been suspected before that this concept might be flawed, see
e.g.~\cite{Sala_JCP93,Dittrich_Pachon_JCP10}, here we provide a simple proof and a
counterexample.

To disprove Lee and Scully's assertion that for energy eigenstates of quantum systems
$\bm{J}\cdot\bm{\nabla}W=0$, note that for eigenstates
$ \bm{\nabla}\cdot \bm{J} = - \partial_tW =0 $, so, with~$\bm{w} = \bm{J}/W$
\begin{equation}
  \label{eq:WignerFlow_VelocityDivergence}
  \bm{\nabla} \cdot \bm{w} = \frac{W\bm{\nabla} \cdot \bm{J} - \bm{J}
    \cdot \bm{\nabla}W}{W^2}  = -\frac{\bm{J}\cdot\bm{\nabla}W}{W^2} \; .
\end{equation}

Therefore Lee and Scully implicitly assume that the flow is Liouvillian which we showed
previously~\cite{Kakofengitis_PRA17} to imply that no quantum terms are present in
Eq.~\eqref{eq:FlowComponents} for~$\bm J$. This is incorrect for the Morse oscillator they
studied~\cite{Lee_Scully_JCP82}.

We confirm our conclusion by a plot of fieldlines of~$\bm{J}$ (to which the velocity
field~$\bm{w}$, where it exists, is tangential) in Fig.~\ref{fig:divergence}~(b). This shows
that Wigner current crosses $W$'s contours, in other words,
$\bm{J} \cdot \bm{\nabla} W \neq 0$.

\subsection{The Non-Crossing Rules do not apply\label{subsec:Appendix_NoCrossing}}

We have shown in Section~\ref{sec_TimeEvol_Lagrangian} that the non-crossing rules by
Trahan and Wyatt are artefacts of the use of the Lagrangian form~(\ref{eq:W_TotalDeriv}).

Daligault also seems to invoke non-crossing rules when he states that for a region~$V_0$
where the Wigner distribution has negative polarity, and $\bm \nabla \cdot \bm{w} < 0$,
\emph{``the trajectories lying in this volume would condense and eventually collapse into
  a volume of zero volume. From the practical viewpoint, the set of initial trajectories
  modelling the whole initial region $V_0$ would eventually describe a tiny
  volume.}''~\cite{Daligault_PRA03}

Lee and Scully's argument~\cite{Lee_Scully_JCP82} for `{Wigner trajectories}', see
section~\ref{App:No_Wigner_Trajectories} above, also amounts to invocation of a
non-crossing rule.

Section~\ref{App:No_Wigner_Trajectories} and Fig.~\ref{fig:divergence}~(b) prove
assertions based on Liouvillian flow and non-crossing rules incorrect. Instead of
trajectories-based on the velocity field~$\bm{w}$, fieldlines of Wigner current~$\bm{J}$
should be used, they are singularity-free and cross zero-contours
of~$W$~\cite{Ole_PRL13,Kakofengitis_EPJP17}.

\subsection{Misconceptions due to incorrect decomposition of the continuity
  equation\label{subsec:Appendix_Errors_Cont_Eq_decomposition}}

In equation~(\ref{eq:FlowComponents}) for $\bm{J}$ the $l=0$-term is the classical force
term rendering the dynamics, if truncated here, Liou\-vil\-lian
($\bm{\nabla} \cdot \bm{w}=0$). The classical form also is \emph{degenerate} in the sense
that the current is zero wherever $W$ is zero~\cite{Kakofengitis_EPJP17}.
\\
In the anharmonic quantum case this is typically not the case, since (lines of) zeros of
the Wigner distribution do not imply that the current stagnates. Instead, this degeneracy
in $\bm{J}$ is lifted due to the quantum terms, of
order~$l \geq 1$~\cite{Kakofengitis_EPJP17}. This implies that zeros of~$W$ are zeros of
$J_x$ but not of~$J_p$. The quantum terms in~\eqref{eq:FlowComponents} shift the lines of
zero of $J_p$ away from those of $J_x$. Only where those lines intersect do stagnation
points of the current exist~\cite{Kakofengitis_EPJP17}, see Fig.~\ref{fig:divergence}~(b)
and~(c). The stagnation points of the current therefore straddle the boundaries of
negative regions of~$W$ where the current gets
inverted~\cite{Ole_PRL13,Kakofengitis_EPJP17}. These stagnation points have special
importance because they are topologically protected~\cite{Ole_PRL13}, and they display
very large local variations of the direction of the current for non-zero values of the
momentum~$p$, a feature alien to classical Hamiltonian flows. These aspects of the
stagnation points of Wigner current were found
recently~\cite{Ole_PRL13,Kakofengitis_EPJP17} although precursors were observed in quantum
\ps studies of Husimi's function~\cite{Skodje_PRA89}.

\paragraph{Incorrect use of Newton trajectories\label{paragraph:Appendix_Newton_Trajectory_Decompositions}}
In reference~\cite{Brosens_SSC10} continuity equation~\eqref{eq:W_Continuity} is
decomposed into its classical term [with~$\bm{v}$ from Eq.~\eqref{eq:Classical_flow}] and
quantum term~$Q$
\begin{equation} 
  \partial_t W + \bm{v} \cdot \bm \nabla W = -\partial_p J_p - \partial_p W \partial_x V = Q\; . 
\label{eq:W_PseudoNewton}
\end{equation}
This is an incorrect decomposition, the correct Lagrangian decomposition is given in
equation~\eqref{eq:W_TotalDeriv}. The authors then formally integrate this equation
propa\-gating their solutions along classical Newtonian
trajectories~\eqref{eq:Trajectory_Solution_Lagrangian}, supposedly fully taking into
account all quantum effects~\cite{Brosens_SSC10,Sels_PASM12}. This is only correct for
quantum-mechanical cases whose hamiltonians have potentials up to second order in
position~$x$~\cite{Takabayasi_PTP54,Royer_FOP92,Brosens_SSC10,Ole_PRL13\anon{,Ole_FreeHOSC_14}},
in which case~$Q=0$. In the anharmonic case, the approach of
references~\cite{Brosens_SSC10,Sels_PASM12} does not allow for the directional
modifications of the current that is so characteristic for quantum dynamics
(see~\cite{Ole_PRL13,Kakofengitis_EPJP17} and Fig.~\ref{fig:divergence}): the Newton trajectory
approach is incorrect.

\paragraph{Incorrect total derivative decomposition\label{paragraph:Appendix_DWDT_Decompositions}}
Carruthers and Zachariasen~\cite{Carruthers_RMP83} decomposed Wigner current according to
$\frac{d W}{dt}=-\partial_p J_p$, the correct expression is
$\frac{d W}{dt}=- W \partial_p w_p$, see equation~(\ref{eq:W_TotalDeriv}).

\paragraph{The $\dot p = \partial_p J_p / \partial_p W $ decomposition error\label{paragraph:Appendix_Error_dpJp_by_dpW}}
Lee~\cite{Lee_PLA90} and Lee and Scully~\cite{Lee_Scully_JCP82,Lee_FoP83,Lee_PRep95}
incorrectly decomposed~$\bm{J}$ using the analogy with classical physics. Imposing the
Liouvillian form
\begin{eqnarray}\label{eq:PseudoClassical_Continuity}
\partial_t W +  \frac{p}{M} \; \partial_x W + \dot p \; \partial_p W  = 0  \; ,
\end{eqnarray}
they concluded that, in the quantum case,~$\dot p = \partial_p J_p / \partial_p W $. Their
formal integration of this equation leads to incorrect results such as those detailed in
Section~\ref{App:No_Wigner_Trajectories}.

This decomposition was criticised by Daligault~\cite{Daligault_PRA03}, criticised and yet
adopted by Sala \emph{et al.}~\cite{Sala_JCP93} and by Henriksen \emph{et
  al.}~\cite{Henriksen_CPL88} (who later concluded though that, based on numerical work,
\emph{``These studies showed a fatal degradation of the distribution
  function''}~\cite{Moller_JPC94}). Decomposition~\eqref{eq:PseudoClassical_Continuity}
was also adopted by, e.g., Muga \emph{et al.}~\cite{Muga_SSC95},
Razavy~\cite{Razavy_PLA96,Razavy_Book_03}, Dias and Prata~\cite{Dias_JMP02}, Zhang and
Zheng~\cite{Zhang_ChPL09}, and reported by Landauer~\cite{Landauer_RMP94}.

\subsection{Can the non-zero divergence of the current be transformed
  away?\label{subsec:Appendix_Error_Daligault_Remove_NonZeroDiv}}

In Reference~\cite{Kakofengitis_PRA17} we established that Wigner current obeys
Liouville's theorem only for systems with potentials at most quadratic in~$x$.

Daligault asks whether one can find a transformation \emph{``that would render
  Hamiltonian} [divergence-free] \emph{quantum fluid dynamics in phase
  space''}~\cite{Daligault_PRA03}. Reference~\cite{Kakofengitis_PRA17} proves this is not
possible. Here we give an additional argument:

The idea of `transforming away' the divergence in $\bm{w}$ amounts to transforming away
the quantum terms in~$\bm{J}$~\cite{Kakofengitis_PRA17} and is ill-conceived: according to
equation~(\ref{eq:W_Solution_Integrated}), a divergence-free velocity field would not
allow for a change of the value of~$W$ along a fieldline of~$\bm{J}$. Since fieldlines
are defined in all of phase space, negativities and quantum coherences could never form.

Based on an analysis different from Daligault's, Sala \emph{et al}.~\cite{Sala_JCP93}
argue that the `Wigner trajectories' of Lee and Scully, see
section~\ref{App:No_Wigner_Trajectories}, exist, that these trajectories follow the
contours of~$W$, and that along them $\frac{dW}{dt}=0$. They modify and reinforce this
statement by saying that \emph{``Liouville's theorem in the form of area preservation of a
  given contour of moving phase points is obeyed as long as the defined contour does not
  touch any of the singularities. The singularities are not only responsible for
  ``destruction" of trajectories. They can also ``create" them."}\cite{Sala_JCP93}

We found here that for anharmonic oscillators $\bm{w}$ is divergence-free only on lines
(of measure zero) in \ps, see Fig.~\ref{fig:divergence}~\figureLabel{c}. In other words,
quantum \ps dynamics of anharmonic systems is non-Liouvillian almost everywhere in \ps. We
have (other than for harmonic oscillator eigenstates and high-temperature thermal states)
not seen evidence of fieldlines of~$\bm{J}$ following $W$'s contours.  The reported
observation of the ``creation or destruction'' of trajectories at a singularity
of~$\bm{w}$ might be due to careless numerics using an adaptive integrator.

\subsection{There are no quantum potentials in \ps\label{subsec:QuantumPotentials}}

The concept of Wigner trajectories was also used for the introduction of the concept of a
``{quantum potential}''~$\tilde V$~\cite{Lee_PLA90} or ``{quantum
  force}''~$\tilde F$~\cite{Razavy_PLA96,Razavy_Book_03}.  The underlying idea is to
identify the term~$\dot p$ in Eq.~\eqref{eq:PseudoClassical_Continuity} with
$\dot p= \tilde F = - \partial_x \tilde V$. Being based on an erroneous identification of
terms in Eq.~\eqref{eq:PseudoClassical_Continuity}, it gives rise to incorrect and
peculiar results such as a force with singularities~\cite{Razavy_PLA96,Razavy_Book_03}.

\subsection{What about positive  \ps distibutions?\label{subsec:PositiveDistributions}}

Proof~\eqref{eq:W_Solution_Integrated} applies to all \ps quantum distributions
with negative values, i.e., to the entire continuum of gaussian smeared distributions of
which Husimi's Q-function is the (positive semidefinite) limit~\cite{Hillery_PR84}.

Certain technical issues are tamed by using positive distributions such as Husimi's
Q-function~\cite{Huaqing_JPCL13} or some thermal Wigner distributions, but we doubt it
changes the fundamental inadmissibility of the trajectory concept for positive
distributions: to use trajectories one would have to find a transformation that removes
the quantum terms in~\eqref{eq:FlowComponents} in order to render the equations first
order in their derivatives and linear in~$W$ (Section~\ref{subsec_nonLinear_w}). But the
quantum terms are always present for anharmonic
systems~\cite{Hillery_PR84,Takahashi_JPSJ86,Veronez_JPA13}.

Particular non-trivial anharmonic systems with special states and special symmetries,
admitting transformations to Lagrangian forms that simplify treatment, might exist. In
general (and in light of failed attempts to find such transformations
for~$W$~\cite{Kakofengitis_PRA17})
suggestions~\cite{PrivateCommunications_and_ReviewersReports} that evolution equations of
non-negative quantum distributions for anharmonic systems might admit trajectory-based
representations appear implausible.

\vspace{-0.25cm}
\section{Conclusions\label{sec_concl}}

Quantum \ps dynamics has frequently been cast into Lagrangian form in order to represent
its transport along trajectories. For anharmonic quantum mechanical systems this leads to
a \ps velocity field~$\bm{w}$ with singularities and singular
divergence~$\bm{\nabla} \cdot \bm{w}$. For anharmonic quantum systems transport-solutions,
using trajectories, are mathematically ill-defined: trajectory-based approaches have to be
avoided.

The occurrence of singularities of~$\bm{\nabla} \cdot \bm{w}$ in the anharmonic quantum
case is \emph{needed and responsible} for the generation of negativities in Wigner's
quantum \ps distribution and thus for the generation of quantum coherences. This
realization provides a deeper understanding of the differences between quantum and
classical \ps dynamics (for which $\bm{\nabla} \cdot \bm{w}=0$). It also explains
the frequently reported poor performance of numerical schemes employing trajectories in
\ps.

Instead of studying quantum \ps dynamics from a Lagrangian trajectory approach it
should primarily be studied from an Eulerian approach centered on Wigner's quantum
\ps current~$\bm{J}$. $\bm{J}$-fieldlines always exist and they reveal intriguing
detail~\cite{Ole_PRL13,Kakofengitis_EPJP17}.

An interesting open question~\cite{PrivateCommunications_and_ReviewersReports} is: how
stable are entangled-trajectories
methods~\cite{Donoso_PRL01,Donoso_JCP03,Dittrich_Pachon_JCP10}?

\vspace{-0.25cm}
\section*{Acknowledgement} 
\newnew{O. S.  feels indebted to Paul Brumer for many discussions about the importance of
  Wigner distribution negativities which spawned these investigations.}



\begin{thebibliography}{67}
\bibitem [{Tra()}]{Trahan_Wyatt_(i)_comment}%
  \BibitemOpen
  \href@noop {} {}\bibinfo {note} {{We cannot comment on statement~(i), being
  unable to confirm it.}}\BibitemShut {Stop}%
\end{thebibliography}
%

\end{document}